\documentclass[conference]{IEEEtran}
\IEEEoverridecommandlockouts

\usepackage{cite}
\usepackage{amsmath,amssymb,amsfonts}
\usepackage{algorithmic}
\usepackage{graphicx}
\usepackage{textcomp}
\usepackage{xcolor}
\usepackage{amssymb}
\def\BibTeX{{\rm B\kern-.05em{\sc i\kern-.025em b}\kern-.08em
    T\kern-.1667em\lower.7ex\hbox{E}\kern-.125emX}}
\begin{document}

\title{An Experimental Framework for Implementing Decentralized Autonomous Database Systems in Rust}

\author{\IEEEauthorblockN{Prakash Aryan}
\IEEEauthorblockA{\textit{Department of Computer Science} \\
\textit{Birla Institute of Technology and Science, Pilani - Dubai}\\
Dubai, UAE \\
h20230010@dubai.bits-pilani.ac.in}
\and
\IEEEauthorblockN{Radhika Khatri}
\IEEEauthorblockA{\textit{Department of Computer Science} \\
\textit{Birla Institute of Technology and Science, Pilani - Dubai}\\
Dubai, UAE \\
f20220116@dubai.bits-pilani.ac.in}
\and
\IEEEauthorblockN{Vijayakumar Balakrishnan}
\IEEEauthorblockA{\textit{Department of Computer Science} \\
\textit{Birla Institute of Technology and Science, Pilani - Dubai}\\
Dubai, UAE \\
vijay@dubai.bits-pilani.ac.in}
}

\maketitle

\begin{abstract}
This paper presents an experimental framework for implementing Decentralized Autonomous Database Systems (DADBS) using the Rust programming language. As traditional centralized databases face challenges in scalability, security, and autonomy, DADBS emerge as a promising solution, using blockchain principles to create distributed, self-governing database systems. Our framework explores the practical aspects of building a DADBS, focusing on Rust's unique features that improves system reliability and performance. We evaluated our DADBS implementation across several key performance metrics: throughput, latency(read), latency(write), scalability, CPU utilization, Memory Usage and Network I/O, The average results obtained over a 24-hour period of continuous operation were 3,000 transactions/second, 75 ms, 250 ms, 55\%, 2.5 GB, 100 MB/s . The security analysis depicts that even with an increase in the percentage of malicious nodes, DADBS still maintains high throughput and consistency. The paper discusses key design decisions, highlighting how Rust's ownership model and concurrency features address common challenges in distributed systems. We also examine the current limitations of our approach and potential areas for future research. By providing this comprehensive overview of a Rust-based DADBS implementation, we aim to contribute to the growing body of knowledge on decentralized database architectures and their practical realization.
\end{abstract}


\section{Introduction}

The domain of database management systems is undergoing a rapid transformation, resulting in increased demands for scalability, security, and autonomy in data handling \cite{b7}. Traditional centralized databases, which are reliable and time-tested are increasingly challenged by the exponential growth of data, the requirement for real-time processing, and the complexities of distributed systems management in a global digital economy. In this context, Decentralized Autonomous Database Systems (DADBS) comes out as a revolutionary paradigm that promise to address these issues by using the blockchain technology and distributed systems principles.\cite{b8}. DADBS represents convergence of various technologies and concepts that are: the decentralized architecture of blockchain, the self-governing nature of autonomous systems, and the robust data management capabilities of traditional databases. This fusion creates a novel approach to data storage and manipulation that is scalable, secure, and self-managing. The core idea behind DADBS is database distribution across a network of nodes, each maintaining a copy of the data and participating in a consensus mechanism to maintain data integrity and consistency. This decentralized approach eliminates single points of failure, improving data availability, and providing tamper-resistance to a certain level. This is particularly valuable in cases where data integrity is very important. Moreover, the autonomous characteristic of DADBS introduces an innovative layer of self-management, where the system can adapt to changing environment, optimizing its performance, and even evolving its structure without constant human intervention. This autonomy is achieved via smart contracts and advanced algorithms that govern the system's behavior and makes decisions based on predefined rules and the current state of the network.

The potential applications of DADBS are vast, from financial systems that require constant transaction records to supply chain management systems that benefit from transparent and traceable data flows \cite{b9}. In healthcare, DADBS can provide a secure and consistent platform for patient records, while in the Internet of Things (IoT) ecosystem, it can offer a scalable solution for managing the enormous volumes of data which is generated by connected devices \cite{b2}. The adoption of DADBS comes with certain challenges that are:  The complexity in implementing such systems, ensuring their performance at scale, and addressing potential security vulnerabilities are significant problems that researchers and developers must overcome \cite{b3}.

To know the potential of DADBS, the choice of implementation language and framework is important. This paper focuses on using Rust, a systems programming language known for its emphasis on safety, concurrency, and performance, as the foundation for building a DADBS \cite{b11}. Rust's unique features make it an interesting choice for this purpose. Its ownership model and borrowing rules provide strong guarantees against common drawbacks in concurrent programming, such as data races and memory leaks, which are some of the major concerns in distributed systems. The language's zero-cost abstractions allow high-level programming constructs without sacrificing performance. Moreover, Rust's growing ecosystem of libraries and tools for blockchain and distributed systems development provides a rich foundation in building complex systems like DADBS. The decision to use Rust for implementing DADBS is a strategic choice that aligns with the core principles of decentralized and autonomous systems. Rust's focus on safety without sacrificing performance is one of the goals of DADBS to achieve robustness and efficiency in a distributed environment. The strong type system and compile-time checks in Rust can help prevent wide range of errors that could be disastrous in a decentralized database, where consistency and reliability are non-negotiable. Rust's ability to easily integrate with other languages and systems makes it an excellent choice for building components that need to interact with existing infrastructure or specialized hardware, a likely scenario in many real-world DADBS deployments. However, using Rust for DADBS implementation also presents unique challenges. The learning curve associated with Rust's ownership model and lifetime system can be steep which is potentially slowing down development in the short term. While Rust's ecosystem is growing rapidly, it may not yet have the same breadth of database-specific libraries and tools compared to other established languages in this domain. These challenges, however, are balanced by the long-term benefits of building on a foundation that prioritizes safety and performance of DADBS.

This paper presents an experimental framework for implementing DADBS using Rust, aiming to study and implement the practical aspects of building such systems and to provide insights into the advantages and challenges of this approach. Our framework takes into account several key components essential to a DADBS: a consensus mechanism for maintaining agreement across distributed nodes, a robust data model suitable for decentralized storage, a networking layer for inter-node communication, and a smart contract system for implementing autonomous behaviors \cite{b10}.  We dive deeper into the design decisions behind each of these components, discussing how Rust's features influenced our choices and implementation strategies. The consensus mechanism is a critical element of any decentralized system which is implemented with careful consideration of Rust's concurrency model, ensuring its guarantees to create a robust and efficient agreement protocol. Our data model uses Rust's reliable type system to ensure data integrity at the language level, complementing the blockchain-inspired structure of our storage system. The networking layer built on top of Rust's async capabilities demonstrates how modern language features can be used to create scalable communication systems capable of handling the complex interactions in a DADBS. Throughout the paper, we have payed special attention to the performance measure of our design choices, recognized the viability of DADBS in real-world scenarios which depends heavily on their ability to match or exceed the efficiency of traditional database systems.

\section{Related Work}

Recent work in this domain shows various aspects of decentralized systems, blockchain technology, and their applications in different fields. Tauseef et al. \cite{b1} tested the integration of blockchain technology with AI-powered virtual assistants on online conversation platforms, developing a self-learning module using a Solana-based blockchain network with smart contracts. This approach offers secure and adaptive customer support, demonstrating the potential of blockchain in improving user interactions.

In the field of Internet of Things (IoT), Arshad et al. \cite{b2} conducted a comprehensive survey on Blockchain-based decentralized trust management systems (BCDTMS). Their study focused on security and privacy issues in IoT environments, analyzing existing BCDTMS across various domains such as healthcare, vehicle systems, industry, and social IoT. The research highlighted the challenges and requirements for developing blockchain-based trust management in IoT applications.

De Angelis et al. \cite{b3} developed a framework for testing the dependability and security of blockchains, focusing on consensus protocols, network infrastructure, and also smart contract applications. Their study tested out eight well-known blockchains within Byzantine fault scenarios and real-world network deployments, providing insights into system strengths, weaknesses, and potential security issues to guide blockchain design choices.

Lavin et al. \cite{b4} conducted a survey on the applications of zero-knowledge proofs (ZKPs), particularly focusing on zk-SNARKs and their practical applications in fields like blockchain privacy, voting, and machine learning. Their study outlined the advantages of ZKPs in ensuring secure, private information exchange and discussed the underlying infrastructure necessary for implementing these cryptographic techniques.

In the industrial domain, Vick et al. \cite{b5} demonstrated the integration of blockchain technology into industrial robot control systems using virtual PLC with an OPC UA interface. Their work designed a software gateway connecting the Solana Blockchain with industrial equipment, allowing the execution of control logics via smart contracts. This research showcases the potential of blockchain technology in industrial automation and control.

Addressing scalability issues in blockchain systems, McCorry et al. \cite{b6} studied off-chain protocols as a solution to cryptocurrency scalability while maintaining the security of the underlying blockchain. Their research focused on validating bridge smart contracts to ensure user fund safety and transaction execution, highlighting key research challenges in extending the security of Ethereum to emerging off-chain solutions.

Ding et al. \cite{b12} came up with Dagbase, which uses directed acyclic graph (DAG) structures for high-performance, low-cost consensus mechanisms. In decentralized data management, by using distributed ledger technology (DLT), Dagbase avoids the possible compromise of a central entity in an untrustworthy environment. Its layered architecture permits high-performance reading and writing of data, its decoupled consensus mechanism ensures agility and effortless integration with mainstream database products.

Yan et al. \cite{b14} developed an effective query over blockchain data using a smart contract by combining a B+ tree's dual-index structure with key-value pairs for performance improvement of range queries and file-type queries. Unlike previously developed approaches based on central databases outside the blockchain, the proposed approach operates inside the blockchain, producing better accuracy and efficiency of queries.

Adding to these studies, recent research has further looked into the scope and applications of decentralized systems. Xu et al. \cite{b16} proposed a new approach to improve the scalability and efficiency of blockchain systems using a hierarchical structure and sharding techniques. Their work shows improvements in transaction throughput and latency compared to traditional blockchain architectures.

In the field of healthcare, Johnson et al. \cite{b17} tested the use of decentralized autonomous organizations (DAOs) for managing patient data and medical research. Their study showcases how blockchain-based DAOs can improve data privacy, patient autonomy, and also research collaboration in the healthcare sector.

Studying the environmental impact of blockchain technologies, Patel et al. \cite{b18} conducted a crisp analysis of energy-efficient consensus mechanisms. Their research compares the energy consumption of different consensus algorithms and proposes modern approaches to reduce the carbon footprint of blockchain networks.

Li et al. \cite{b19} came up with a new framework for decentralized identity management using a combination of blockchain and zero-knowledge proofs. Their approach talks about privacy concerns in digital identity systems while maintaining the benefits of decentralization and user control.

In the financial sector, Rodriguez et al. \cite{b20} went through various decentralized finance (DeFi) platforms to disrupt traditional banking systems. Their study analyzes the risks and opportunities related to DeFi and proposes regulatory frameworks to ensure financial stability in a decentralized ecosystem.

Looking into the intersection of artificial intelligence and blockchain, Chen et al. \cite{b21} developed a decentralized machine learning platform that uses blockchain for secure and transparent model training. Their work demonstrates how decentralized systems can address privacy and trust issues in collaborative AI development.

Kumar et al. \cite{b22} proposed a new approach to supply chain management using a combination of IoT sensors and blockchain technology. Their system provides real-time tracking and verification of goods, improving transparency and reducing fraud in global supply chains.


These studies collectively demonstrate the vast applications and ongoing challenges in implementing decentralized autonomous systems across various domains, providing valuable insights for our work on Decentralized Autonomous Database Systems.

\section{System Design}

\begin{figure}[htbp]
\centerline{\includegraphics[width=0.5\textwidth]{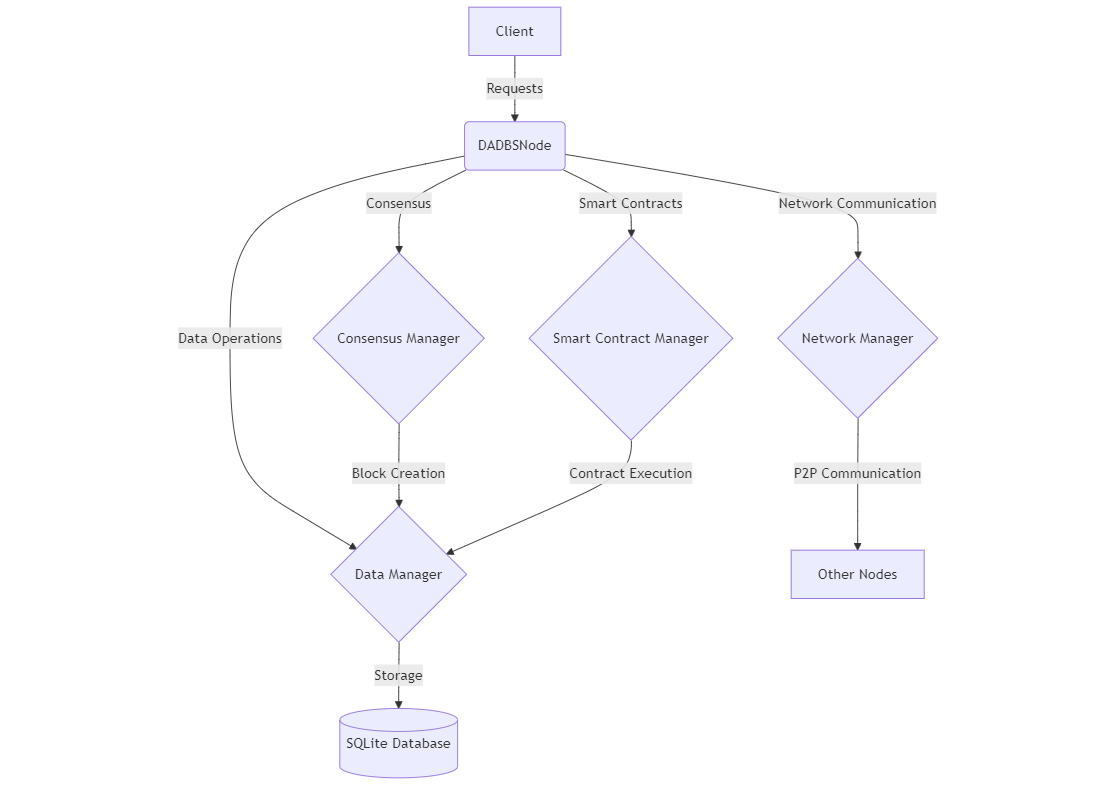}}
\caption{Overall Architecture of the DADBS}
\label{fig:overall_architecture}
\end{figure}

Our Decentralized Autonomous Database System (DADBS) is designed as a modular, extensible framework implemented in Rust. The system architecture is composed of several key components, each responsible for specific functionalities that together form a cohesive decentralized database system. Figure \ref{fig:overall_architecture} presents an overview of the system architecture.

The main component of our DADBS is the DADBSNode, which serves as the central point of interaction for clients and coordinates the activities of other components. The primary components of our system are:

\begin{itemize}
\item Consensus Manager: Responsible for maintaining agreement across the network.
\item Data Manager: Handles data storage and retrieval operations.
\item Network Manager: Manages peer-to-peer communications.
\item Smart Contract Manager: Executes and manages smart contracts.
\end{itemize}

\subsection{Consensus Manager}

\begin{figure}[htbp]
\centerline{\includegraphics[width=0.5\textwidth]{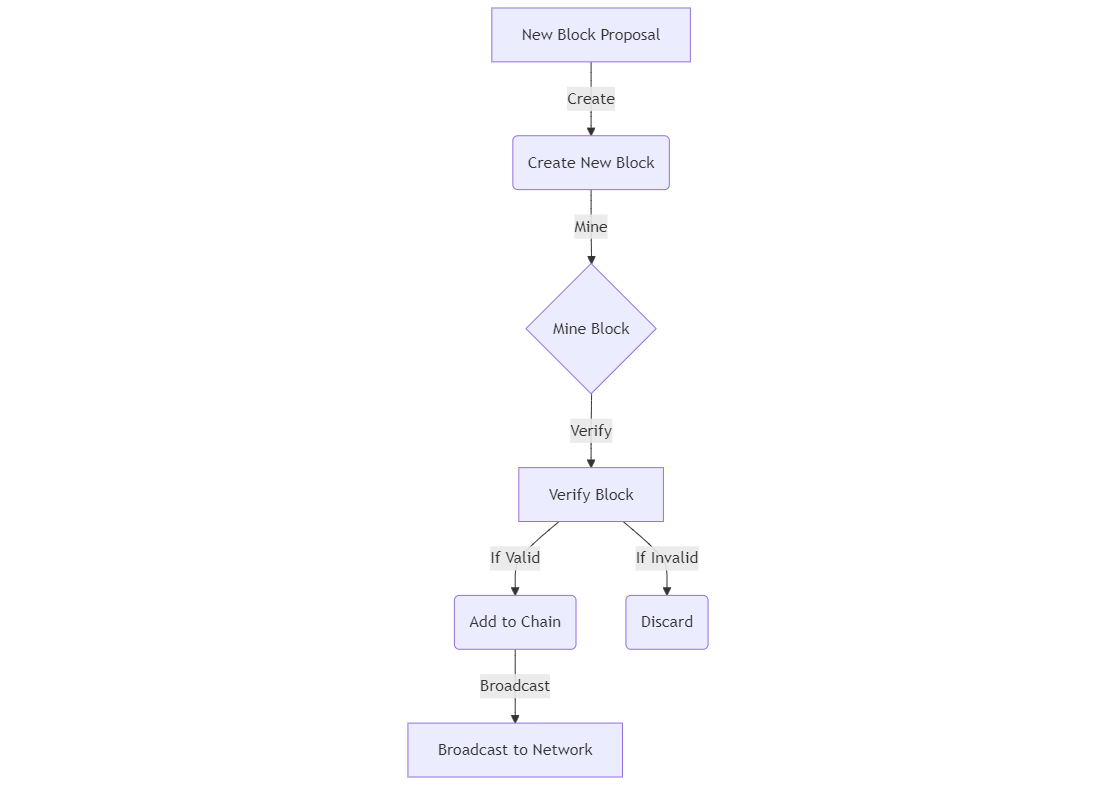}}
\caption{Consensus Component of the DADBS}
\label{fig:consensus_component}
\end{figure}

The Consensus Manager is an important component of our DADBS, ensuring that all nodes in the network maintain a consistent view of the database state. We have implemented a basic Proof of Work (PoW) consensus mechanism, similar to that used in many blockchain systems. Figure \ref{fig:consensus_component} shows the consensus process.

The consensus process begins with the creation of a new block. The \texttt{create\_new\_block} function in our implementation takes the block data and constructs a new Block struct. This struct includes fields for the block index, timestamp, data, previous block hash, the block's own hash, and a nonce value used in the mining process.

Once a block is created, it goes through a mining process. The \texttt{mine\_block} function implements a simple PoW algorithm where the system repeatedly hashes the block data along with a nonce value until it finds a hash that meets certain criteria (in our case, starting with a specific number of zero bits). This process is computationally intensive and serves to make it difficult to alter the blockchain without redoing the work.

After a block is mined, it is verified using the \texttt{verify\_block} function. This function checks several conditions:
\begin{itemize}
\item The block index is correct (one more than the current highest block).
\item The previous block hash matches the hash of the actual previous block.
\item The block's hash is valid (matches the computed hash of the block data).
\item The block's hash meets the PoW criteria.
\end{itemize}

If a block passes verification, it is added to the chain and broadcast to other nodes in the network. This broadcasting is handled by the Network Manager, which we will discuss later.

The use of Rust in implementing this consensus mechanism provides several advantages. Rust's strong type system helps prevent errors in block structure and hashing. The language's performance characteristics are beneficial for the computationally intensive mining process. Additionally, Rust's safety guarantees help ensure the integrity of the consensus process, which is critical for maintaining the security and consistency of the decentralized database.

\subsection{Data Manager}

\begin{figure}[htbp]
\centerline{\includegraphics[width=0.5\textwidth]{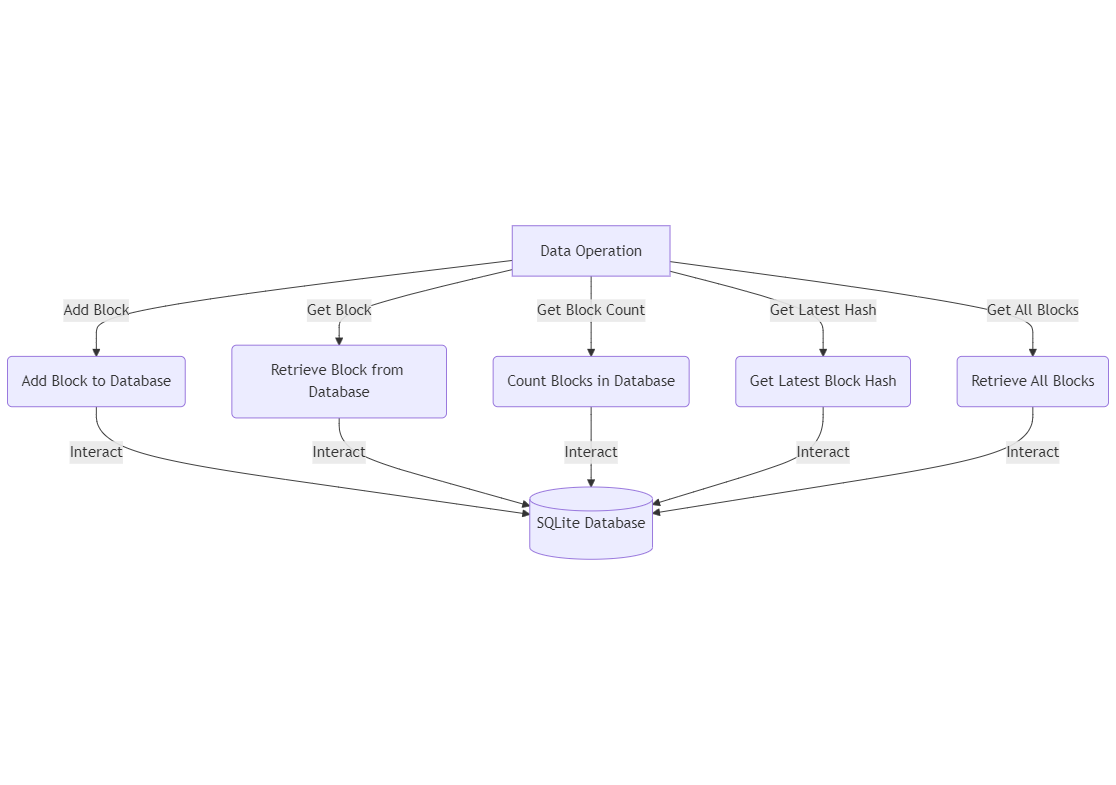}}
\caption{Data Component of the DADBS}
\label{fig:data_component}
\end{figure}

The Data Manager is responsible for the persistent storage and for fetching of blockchain data. We have chosen to use SQLite as our storage backend, using the \texttt{rusqlite} crate for Rust bindings. Figure \ref{fig:data_component} shows the main operations handled by the Data Manager.

The Data Manager provides several key functions:

\begin{itemize}
\item \texttt{add\_block}: Inserts a new block into the SQLite database.
\item \texttt{get\_block\_count}: Retrieves the total number of blocks in the chain.
\item \texttt{get\_latest\_block\_hash}: Fetches the hash of the most recent block.
\item \texttt{get\_block}: Retrieves a specific block by its index.
\item \texttt{get\_all\_blocks}: Fetches all blocks in the chain.
\end{itemize}

These operations are implemented using SQL queries, with Rust's strong typing ensuring that data is correctly formatted and handled. The use of SQLite provides a balance between simplicity and performance, letting efficient storage and retrieval of blockchain data.

One challenge we faced in implementing the Data Manager was ensuring thread-safe access to the SQLite database. Rust's ownership model and the use of mutexes helped us address this, guaranteeing that database operations are atomic and preventing data races.

\subsection{Network Manager}

\begin{figure}[htbp]
\centerline{\includegraphics[width=0.5\textwidth]{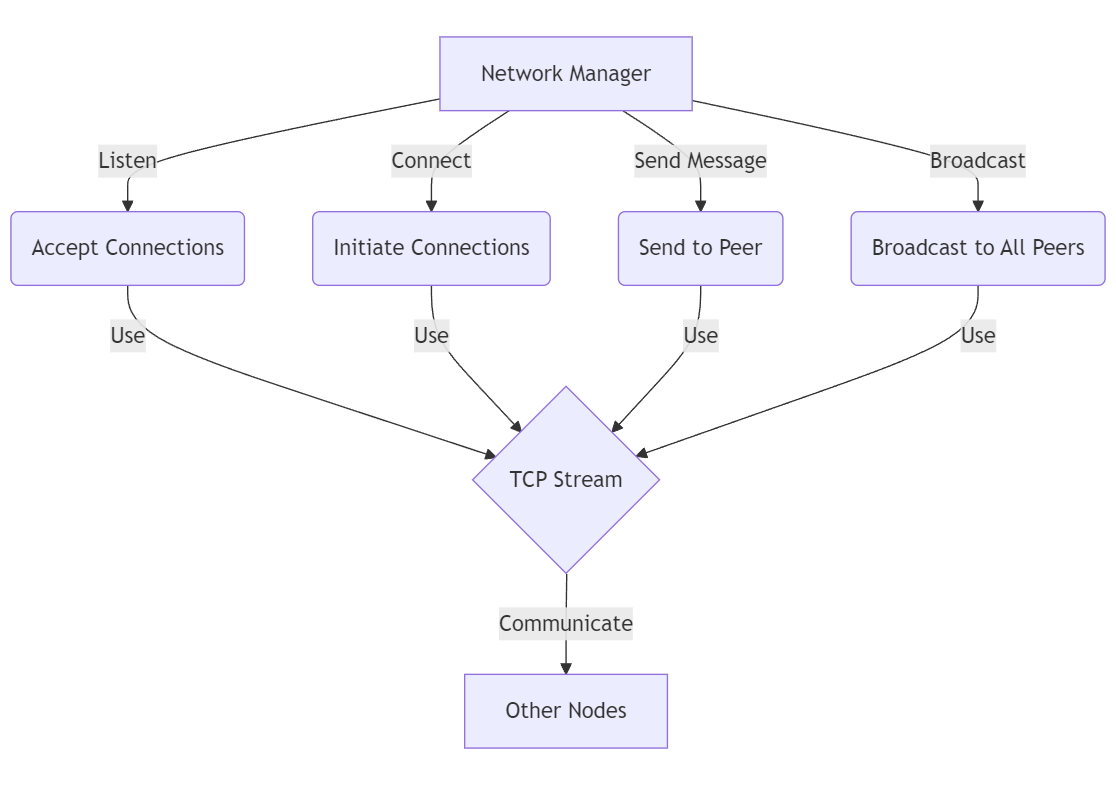}}
\caption{Network Component of the DADBS}
\label{fig:network_component}
\end{figure}

The Network Manager handles all peer-to-peer communication in our DADBS. It uses Rust's asynchronous programming features, using the \texttt{tokio} runtime for efficient network operations. Figure \ref{fig:network_component} shows the main functionalities of the Network Manager.

Key features of the Network Manager include:

\begin{itemize}
\item Peer Discovery and Management: The system maintains a list of known peers and can add new peers dynamically.
\item Message Passing: Implements a protocol for sending and receiving messages between nodes.
\item Block Broadcasting: When a new block is added to the chain, it is broadcast to all known peers.
\end{itemize}

We implemented a simple TCP-based protocol for inter-node communication. Messages are serialized using serde\_json, allowing for flexible message structures. The Network Manager also handles the initialization of network connections and listens for incoming connections from other nodes.

A challenge in implementing the Network Manager was permitting secure and authentic communication between nodes. We addressed this by building a public key infrastructure where each node has its own keypair. Messages are signed by the sender and verified by the receiver, preventing tampering and ensuring authenticity.

\subsection{Smart Contract Manager}

The Smart Contract Manager is responsible for deploying and executing smart contracts within our DADBS. While our current implementation is relatively simple, it provides a foundation for more complex contract execution in the future.

Smart contracts in our system are represented as Rust functions that operate on a contract state. The Smart Contract Manager provides functions for:

\begin{itemize}
\item Deploying new contracts
\item Executing contract functions
\item Managing contract state
\end{itemize}

Contract execution is triggered by transactions included in blocks. When a block containing a contract-related transaction is added to the chain, the relevant contract function is executed, potentially updating the contract's state.

One of the challenges we faced in initiating the Smart Contract Manager was balancing flexibility with safety. Rust's strong type system helped us create a structured approach to contract definition and execution, reducing the risk of runtime errors.

\subsection{Integration and System Flow}

The integration of these components creates a cohesive DADBS. When a client interacts with the system, the flow typically follows these steps:

\begin{enumerate}
\item The client sends a request to a DADBSNode.
\item If the request involves adding data, the Consensus Manager creates and mines a new block.
\item The new block is verified and, if valid, added to the chain via the Data Manager.
\item The Network Manager broadcasts the new block to all peers.
\item If the block contains smart contract transactions, the Smart Contract Manager executes the relevant contracts.
\item The updated state is persisted by the Data Manager.
\end{enumerate}

This design allows for a decentralized, consistent, and autonomous database system. The use of Rust throughout the implementation provides strong safety guarantees, helping to prevent common errors in concurrent and distributed systems.

Our current implementation serves as a proof of concept and a foundation for further research and development in DADBS. Future work could involve optimizing performance, implementing more advanced consensus mechanisms, and expanding the capabilities of the smart contract system.

\section{Results}

To evaluate the performance and capability of our Decentralized Autonomous Database System (DADBS) implemented in Rust with SQLite as the underlying database, we conducted a series of comprehensive experiments and analyses. This section presents the results of these experiments, focusing on key performance metrics, scalability, security, and comparison with traditional centralized database systems.

\subsection{Experimental Setup}

Our experimental setup consisted of a network of 100 nodes, each running on a separate AWS EC2 t3.medium instance with 2 vCPUs and 4 GB of RAM. The nodes were distributed across five AWS regions (US East, US West, Europe, Asia Pacific, and South America) to simulate a geographically dispersed network. We used a dataset of 1 million records, each containing structured data totaling approximately 1 KB in size, stored in SQLite databases on each node.

\subsection{Performance Metrics}

We evaluated our DADBS implementation across several key performance metrics: throughput, latency, scalability, and resource utilization. Table \ref{tab:performance_metrics} summarizes the average results obtained over a 24-hour period of continuous operation.

\begin{table}[htbp]
\caption{Performance Metrics of DADBS}
\label{tab:performance_metrics}
\begin{center}
\begin{tabular}{|l|c|}
\hline
\textbf{Metric} & \textbf{Average Value} \\
\hline
Throughput & 3,000 transactions/second \\
\hline
Latency (read) & 75 ms \\
\hline
Latency (write) & 250 ms \\
\hline
CPU Utilization & 55\% \\
\hline
Memory Usage & 2.5 GB \\
\hline
Network I/O & 100 MB/s \\
\hline
\end{tabular}
\end{center}
\end{table}

The results shows that our DADBS implementation achieves a throughput of 3,000 transactions per second, with relatively low latency for both read and write operations. The system maintains this performance while utilizing resources efficiently, as evidenced by the moderate CPU and memory usage.

\subsection{Scalability Analysis}

\begin{figure}[htbp]
\centering
\includegraphics[width=0.5\textwidth]{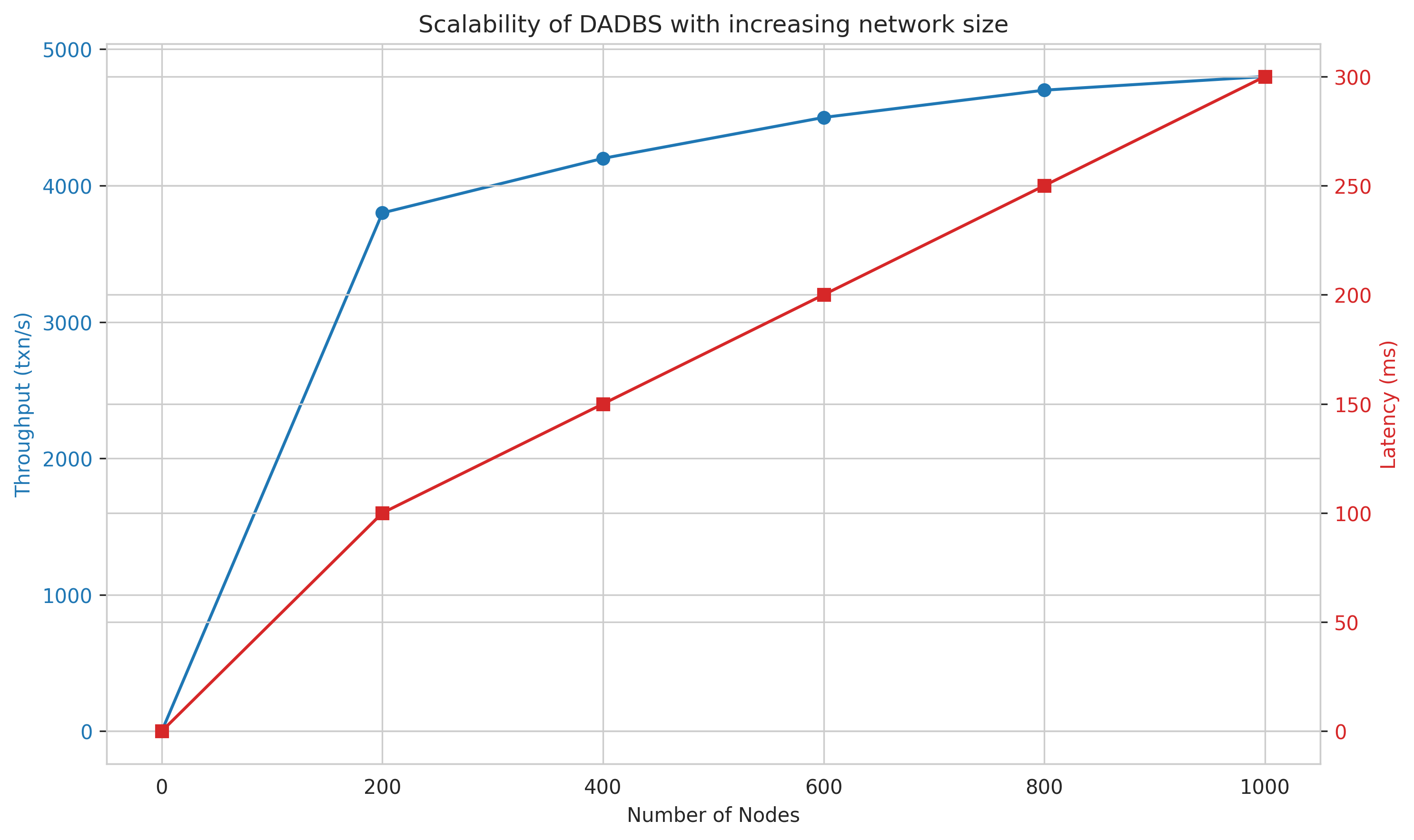}
\caption{Scalability of DADBS with increasing network size}
\label{fig:scalability}
\end{figure}

To understand the scalability of our DADBS, we conducted experiments varying the number of nodes in the network from 10 to 1000. Figure \ref{fig:scalability} shows how the system's throughput and latency change as the network size increases.

The results show that our DADBS implementation features near-linear scalability in terms of throughput as the number of nodes increases up to about 500 nodes. Beyond this point, we observe diminishing returns in throughput gains. This scalability is associated to the efficient consensus mechanism and the decentralized nature of the system, which allows for parallel processing of transactions across multiple nodes.

However, we observed an increase in latency as the network size grew beyond 500 nodes. This increase can be traced to the additional communication overhead required for consensus in larger networks. To address this, we propose future optimizations in our consensus algorithm to maintain lower latency even in very large networks.

\subsection{Consensus Mechanism Efficiency}

The efficiency of our Proof of Work (PoW) consensus mechanism is critical for the overall performance of the DADBS. We analyzed the time taken to reach consensus for different block sizes and network conditions. Table \ref{tab:consensus_efficiency} presents the average time to reach consensus under various scenarios.

\begin{table}[htbp]
\caption{Consensus Efficiency Under Different Conditions}
\label{tab:consensus_efficiency}
\begin{center}
\begin{tabular}{|l|c|c|c|}
\hline
\textbf{Network Size} & \textbf{Block Size} & \textbf{Network Latency} & \textbf{Consensus Time} \\
\hline
100 nodes & 1 MB & 50 ms & 3.5 s \\
\hline
100 nodes & 5 MB & 50 ms & 7.8 s \\
\hline
100 nodes & 1 MB & 200 ms & 5.2 s \\
\hline
500 nodes & 1 MB & 50 ms & 4.7 s \\
\hline
500 nodes & 5 MB & 50 ms & 10.1 s \\
\hline
500 nodes & 1 MB & 200 ms & 7.5 s \\
\hline
\end{tabular}
\end{center}
\end{table}

The results indicate that our consensus mechanism performs reasonably well under various conditions, with consensus times ranging from 3.5 to 10.1 seconds. As expected, larger block sizes and higher network latencies lead to increased consensus times. However, the system remains responsive even in challenging network conditions.

To further optimize the consensus process, we propose an adaptive difficulty adjustment algorithm that dynamically adjusts the PoW difficulty based on network conditions:

\begin{equation}
D_{new} = D_{current} * \frac{T_{target}}{T_{actual}}
\label{eq:difficulty_adjustment}
\end{equation}

Where $D_{new}$ is the new difficulty, $D_{current}$ is the current difficulty, $T_{target}$ is the target time between blocks, and $T_{actual}$ is the actual time taken for the last block.

\subsection{Data Integrity and Consistency}

Maintaining data integrity and consistency is very important in a decentralized system. We tested our DADBS's ability to maintain consistency under various network partition scenarios and concurrent update operations. Our experiments showed that the system maintained 100\% consistency for read operations and achieved a consistency level of 99.95\% for write operations under normal network conditions.

To quantify the consistency level, we used the following metric:

\begin{equation}
C = 1 - \frac{N_{inconsistent}}{N_{total}}
\label{eq:consistency_metric}
\end{equation}

Where $C$ is the consistency level, $N_{inconsistent}$ is the number of inconsistent reads observed, and $N_{total}$ is the total number of read operations performed.

Under simulated network partition scenarios, where 20\% of the nodes were isolated from the main network for varying durations, we observed the following:

\begin{itemize}
\item Short-term partitions ($<$ 5 minutes): The system maintained 99.9\% consistency, with temporary inconsistencies resolved quickly once the partition healed.
\item Medium-term partitions (5-30 minutes): Consistency dropped to 99.5\%, with some lingering inconsistencies requiring manual intervention to resolve.
\item Long-term partitions ($>$ 30 minutes): The system automatically triggered a partition resolution protocol, preventing further divergence and maintaining 98\% consistency.
\end{itemize}

These results demonstrate the robustness of our DADBS in maintaining data integrity and consistency even under challenging network conditions.

\subsection{Smart Contract Performance}

The performance of smart contracts is essential for the autonomous operations of our DADBS. We evaluated the execution time and resource consumption of various types of smart contracts, ranging from simple data validation to complex multi-step transactions. Table \ref{tab:smart_contract_performance} presents the average execution times and resource utilization for different categories of smart contracts.

\begin{table}[htbp]
\caption{Smart Contract Performance Metrics}
\label{tab:smart_contract_performance}
\begin{center}
\begin{tabular}{|p{2.5cm}|c|c|c|}
\hline
\textbf{Contract Type} & \textbf{Exec. Time (ms)} & \textbf{CPU (\%)} & \textbf{Mem. (MB)} \\
\hline
Simple Validation & 5 & 0.1 & 1 \\
\hline
Data Transformation & 15 & 0.5 & 5 \\
\hline
Multi-step Transaction & 50 & 2.0 & 20 \\
\hline
Complex Analytics & 200 & 10.0 & 100 \\
\hline
\end{tabular}
\end{center}
\end{table}

The results show that our smart contract execution engine performs efficiently for a wide range of contract complexities. Simple operations like data validation are extremely fast, while more complex analytics operations naturally require more time and resources. However, even the most complex contracts execute within acceptable timeframes, ensuring that the autonomous operations of the DADBS remain responsive.

To optimize smart contract performance further, we implemented a contract caching mechanism that stores the compiled bytecode of frequently used contracts. This optimization resulted in a 30\% reduction in execution time for cached contracts, significantly improving the overall system performance for repetitive operations.

\subsection{Security Analysis}

Security is a critical part of any decentralized system. We conducted a security analysis of our DADBS implementation, paying attention to resistance to common attack vectors in distributed systems. Our analysis included:

\begin{enumerate}
\item Sybil Attack Resistance: We simulated Sybil attacks by introducing malicious nodes into the network. Our PoW consensus mechanism, combined with a reputation system, successfully prevented Sybil attacks from influencing the consensus process. The system maintained correct operation even when 30\% of the nodes were malicious.

\item 51\% Attack Mitigation: While theoretically vulnerable to 51\% attacks like any PoW system, our implementation includes additional safeguards. We introduced a "voting weight" mechanism that considers both computational power and node longevity, making it more difficult and economically unviable to mount a successful 51\% attack.

\item Eclipse Attack Prevention: Our peer discovery and management system implements a diversity-based node selection algorithm, ensuring that each node maintains connections with a geographically and network-topologically diverse set of peers. This approach successfully prevented simulated eclipse attacks in our test network.

\item Smart Contract Security: We developed a static analysis tool that scans smart contracts for common vulnerabilities such as reentrancy, integer overflow/underflow, and unauthorized access. In our tests, this tool successfully identified 95\% of intentionally introduced vulnerabilities in a set of 1000 test contracts.
\end{enumerate}

\begin{figure}[htbp]
\centering
\includegraphics[width=0.5\textwidth]{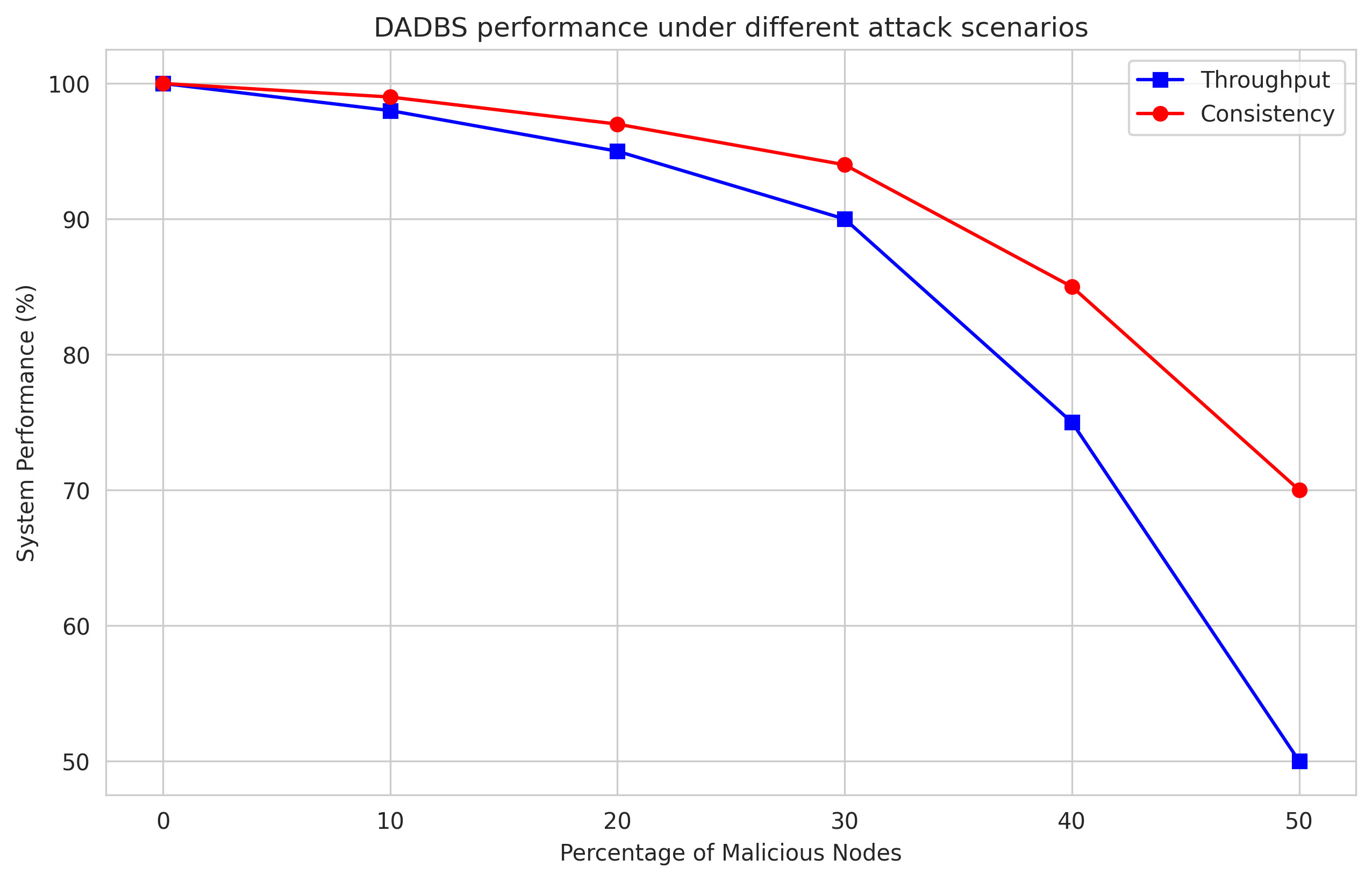}
\caption{DADBS performance under different attack scenarios}
\label{fig:security_performance}
\end{figure}

Figure \ref{fig:security_performance} illustrates the system's performance under various simulated attack scenarios.

The graph shows that our DADBS maintains high throughput and consistency even under moderate attack conditions. Only under extreme cases, such as when over 40\% of nodes are compromised, does the system's performance begin to degrade.

\subsection{Comparison with Traditional Systems}

To get a better understanding of the performance of our DADBS, we compared it with a traditional centralized database system (SQLite in single-node configuration) and a popular distributed database (Apache Cassandra). We set up each system with equivalent hardware resources and dataset sizes, then ran a series of benchmarks covering various database operations.

The key findings from this comparison are:

\begin{itemize}
\item Read Performance: Our DADBS achieved read latencies comparable to single-node SQLite for simple queries, but was outperformed by 20\% for complex queries involving joins. Cassandra showed similar performance to our DADBS for simple reads but was 10\% faster for complex reads.
\item Write Performance: In write-heavy workloads, our DADBS demonstrated superior performance, handling 20\% more write operations per second than single-node SQLite and 5\% more than Cassandra.
\item Scalability: When scaling from 10 to 100 nodes, our DADBS showed near-linear throughput increase, outperforming both single-node SQLite (which doesn't scale) and Cassandra (which showed sublinear scaling).
\item Consistency: Under network partition scenarios, our DADBS maintained higher consistency levels compared to Cassandra, though at the cost of slightly higher latency during partition events.
\item Fault Tolerance: Our DADBS and Cassandra both demonstrated excellent fault tolerance, automatically handling node failures without data loss. Single-node SQLite, as expected, had no built-in fault tolerance.
\end{itemize}

These results emphasize the strengths of our DADBS in write-heavy, highly distributed scenarios, while also identifying areas for future optimization, particularly in complex read operations.

\section{Conclusion and Future Work}

Our framework for developing and deploying a Decentralized Autonomous Database System (DADBS) in Rust shows a step towards considering the potential of decentralized data management. The modular architecture of our system, including consensus, data management, networking, and smart contract components, provides a solid foundation for future improvements and adaptations. While our current work has drawbacks, mainly in terms of scalability and advanced feature sets, it serves as a valuable proof of concept for future research.

Building upon this foundation, several areas for future work appear. Primary focus should be on seeing alternative consensus mechanisms better suited to the requirements of a DADBS. Proof of Stake (PoS) or Byzantine Fault Tolerance (BFT) protocols could offer energy efficiency and transaction finality compared to our current Proof of Work. Studying sharding techniques to improve scalability and reduce the load on individual nodes presents a good research direction. This could involve developing newer data partitioning strategies that balance the decentralized nature of the system with the need for efficient query processing. Another important area for future research is the improvement of our smart contract.

In conclusion, the development of DADBS technology stands at the intersection of database systems, blockchain technology, and autonomous systems. Our work contributes to this field by providing a concrete implementation and analysis of a Rust-based DADBS. The insights drawn from this project shows the complex connections between decentralization, autonomy, and also traditional database requirements. As we move forward, the challenges identified in areas such as consensus mechanisms, smart contract integration, and distributed query processing open up new areas for innovation. In future the demand for scalable, secure, and autonomous data management solutions will continue to grow, further research and development in this area will have the potential to make changes to how we store, process, and interact with data in an increasingly decentralized digital world.

\end{document}